\documentclass[10pt]{article}
\usepackage{graphicx}
\usepackage{rotating} 
\begin{document}

\title{Thera als Angelpunkt  der \"agyptischen und israelitischen Chronologien} 
\author{ W. Woelfli\footnote{ Institute for Particle Physics, ETHZ H\"onggerberg, CH-8093 Z\"urich, Switzerland (Prof.\,emerit.); e-mail: woelfli@phys.ethz.ch .} }

\maketitle

\centerline{\bf Abstract}{\noindent
\emph{The Exodus was causally connected to the eruption of Thera. The time of this geological event is determined by  C14 dating as 1627--1600 BC. The chaos that followed this catastrophe led to upheavals which distorted the chronologies. Taking periods of foreign domination into account leads to an agreement between the chronologies of Egypt and Israel.}

\centerline{\bf Zusammenfassung}{\noindent
\emph{Der Exodus wurde durch den Ausbruch von Thera erm\"oglicht, welcher als geologisches Ereignis durch C14 Datierungen auf die Zeit 1627--1600 BC festgelegt ist. Das durch die Katastrophe bewirkte Chaos f\"uhrte zu Umw\"alzungen, welche die Chronologien verzerrten. Die Ber\"ucksichtigung der Zeiten der Fremdherrschaft f\"uhrt zu einer \"Ubereinstimmung der \"agyptischen und israelischen Chronologien.}

\section{Einf\"uhrung}	
Im Folgenden werden wir die bisher bekannten Zeitangaben f\"ur die \"agyptischen Dynastien und Pharaohs so auflisten, dass sie mit den im Alten Testament erw\"ahnten Zeiten f\"ur die jud\"aischen Richter und K\"onige verglichen werden k\"onnen (Fig.\,1). F\"ur die Darstellung der \"agyptischen Chronologie verwenden wir prim\"ar das Buch von Peter S. Clayton, Chronicle of the Pharaohs [1]. F\"ur die Darstellung der j\"udischen Chronologie ben\"utzen wir vorwiegend  das Alte Testament nach der deutschen \"Ubersetzung D. Martin Luthers [2] sowie das von Jack Finegan verfasste Handbook of Biblical Chronology [3]. In letzterem wird ausf\"uhrlich die Chronologie des Eusebius diskutiert, die wir als Basis f\"ur unsere Betrachtungen der j\"udischen Geschichte ben\"utzen werden. Manetho [4] und Flavius Josephus [5] sind erg\"anzende Quellen.
\section{Chronologie der \"agyptischen Geschichte}
Die \"Agyptologen unterteilen die alt\"agyptische Geschichte in das Alte, das Mittlere  und das Neue Reich (Fig.\,1). Der als Cambridge Ancient History Record  (CAHR) bezeichnete Zeitmassstab basiert auf den historisch mehr oder weniger gut belegten Regierungsdauern der verschiedenen Dynastien. Zwischen diesen drei Zeitabschnitten gibt es jedoch drei Interregnumsphasen, deren zeitliche Abfolgen und Dauer nicht gesichert sind. Ihnen gilt unsere volle Aufmerksamkeit.\vspace{0.3cm}

 3150 --   2686 BC:  Fr\"uhdynastische Periode
 
 2686 -- 2181 BC: Altes Reich \vspace{0.3cm}
 
Ab etwa 3150 BC begann die fr\"uhdynastische Periode. Sie dauerte rund 464 Jahre. Das Alte Reich begann ca.\,um 2686 BC mit der 3.\,Dynastie und endete um 2181 BC mit der 6.\,Dynastie. Die zahlreichen Pyramidenbauten zeugen von einer gewaltigen Schaffenskraft. Allf\"allige Zweifel \"uber die relativ kurze Bauzeit f\"ur all diese Monumente konnten im Rahmen einer systematischen Datierung von 25 Monumenten  aus dieser Zeitspanne mit der C14-Methode entkr\"aftet werden [6]. Der Vergleich zwischen den kalibrierten C14--Bereichen und der historischen Chronologie  von Clayton [1]  zeigt f\"ur die im Bereich des Alten Reichs gemessenen  21 Objekte eine systematische Verschiebung von ca.\,200 Jahre. Das gemessene Alter der 9 aus dem Mittleren Reich stammenden Proben stimmt innerhalb der Fehlergrenzen gut mit der Historischen  Chronologie von  Clayton \"uberein.

\subsection{2181 -- 2040 BC: Erste Zwischen Periode}
Gegen Ende der 6.\,Dynastie nahmen die inneren Wirren zu und das Alte Reich zerfiel aus unbekannten Gr\"unden. Manetho [4] erw\"ahnt z.B.\,eine 7.\,Dynastie. die angeblich 70 K\"onige umfasste, insgesamt aber nur w\"ahrend 70 Tage regierte. Etwas sicherer ist die Existenz der 8.\,Dynastie. Sie bestand vermutlich aus 17  K\"onigen, die insgesamt aber nur 20 Jahre regierten. Zwischen den lokalen F\"ursten der 9.\,und 10.\,Dynastien herrschte ein st\"andiger Machtkampf um die Herrschaft \"uber \"Agypten. Diese erste Interregnumsphase dauerte etwa von 2181 bis 2040 BC.

\subsection{2040 -- 1782 BC: Mittleres Reich}
Das Mittlere Reich startete um 2040 BC mit einer Wiedervereinigung unter dem 4.\,K\"onig der 11.\,Dynastie. Die Dynastie selber begann mit einer Serie von 3 K\"onigen in Serie vorab. In den K\"onigslisten von Saqqara und Abydos wird Mentuhotep II  als letzter K\"onig der 11.\,Dynastie erw\"ahnt. Nach Manetho bestand die 12.\,Dynastie aus 7 K\"onigen aus Theben. Der erste K\"onig dieser Dynastie war Amenemhet I. Er regierte fast 30 Jahre. Seine lange Regierungszeit brachte  die l\"angst ersehnte politische und wirtschaftliche Stabilit\"at f\"ur Ganz \"Agypten. Sie dauerte fast 200 Jahre. Amenemhet III war der letzte grosse Regent im Mittleren Reich. Mit ihm endete nicht  nur seine Dynastie, sondern um 1782 BC auch die als Mittleres Reich bezeichnete Struktur. C14--Datierungen an Proben von Pyramiden  aus der 8.\,und  12.\,Dynastie best\"atigen die historische Chronologie [6].

\subsection{1782 -- 1570 BC:  Zweite Zwischen Periode}
Der \"Ubergang zur 13.\,Dynastie geschah offenbar friedlicher als urspr\"unglich vermutet. Die neue Dynastie umfasste zuerst 10 K\"onige welche 70 Jahre regierten. Gegen Ende der 13.\,Dynastie etablierte sich eine obskure 14.\,Dynastie, die vom \"ostlichen Teil des Nildeltas aus das  Zepter \"ubernahm. Sie regierte angeblich das gesamte Land w\"ahrend rund 57 Jahre. Die verg\"anglichen (kurzlebigen) K\"onige der 14.\,Dynastie waren nicht die einzigen, die sich im Machtbereich der 13.\,Dynastie etablierten. Eine Serie von semitischen K\"onigen breitete sich in der \"ostlichen W\"uste und im \"ostlichen Bereich des Nildeltas aus. Der Hauptharst, die so genannten Hyksos formierten schliesslich die 15.\,und 16.\,Dynastie. W\"ahrend der Macht\"ubernahme wurde ca.\,um 1720 BC Memphis zerst\"ort. \"Uber die endg\"ultige Eroberung des Mittleren Reiches durch die Hyksos wird ausf\"uhrlich von Manetho [4] und Flavius Josephus [5], einem j\"udischen Historiker, mit unterschiedlichen Schlussfolgerungen berichtet. W\"ahrend die Hyksos die Kontrolle \"uber den n\"ordlichen Teil von \"Agypten gewannen, mit Avaris als Residenzstadt, formierte sich in Theben, dem s\"udlichen Teil von \"Agypten, die 17.\,Dynastie. Einige der ersten K\"onige sind bekannt unter dem Namen Intef. Die Feindseligkeiten eskalierten w\"ahrend der Regierungszeit des  Seqenenre Tao, im Kampf gefallen, und seinem Sohn Kamose, der den Befreiungskampf fortsetzte. Die vollst\"andige R\"uckeroberung des n\"ordlichen Teils von \"Agypten gelang schliesslich nach zahl\-rei\-chen K\"ampfen um ca.\,1570 BC. Damit begann das  Neue K\"onigreich mit der ber\"uhmten 18.\,Dynastie und Ahmose als ihr erster K\"onig. 

\subsection{1570 -- 1070 BC: Das Neue K\"onigreich}
Das Neue Reich bestand aus 3 Dynastien. Die 18.\,Dynastie dauerte von 1570 bis 1546 BC und umfasste insgesamt 14 K\"onige. 8 K\"onige bildeten die 19.\,Dynastie. Sie  regierten von 1293 bis 1185 BC. Die 20.\,Dynastie bestand aus 10 K\"onigen, die von 1185 bis 1070 BC herrschten.
Die Taten, bezw.\,die Untaten dieser Despoten werden im Buch von Peter A. Clayton [1] fein s\"auberlich aufgelistet und diskutiert. Wir werden darauf zur\"uckkommen, sobald wir die j\"udische Chronologie behandelt haben.  

\subsection{1069 -- 525 BC: Die Dritte Zwischen Periode}
In der Dritten Zwischen Periode regierten  die folgenden Machthaber:
\begin{tabbing}
soviele regierende Hohe Priester \= Theben nochmals Theben\=    \kill
\, 9\,\, regierende Hohe Priester,    \>           Theben, 1080 -- 945 BC \\
21. Dynastie mit 7 K\"onigen,   \>          Tanis, 1069 -- 945 BC   \\
22. Dynastie mit 11 K\"onigen,   \>        Tanis, 945  -- 715 BC   \\
23. Dynastie mit 6 K\"onigen,     \>        Lybian Anarchy, 818 -- 715 BC  \\
24. Dynastie mit 2 K\"onigen,       \>       Sais, 727 -- 715 BC  \\
25. Dynastie mit 5 K\"onigen,        \>      Nubian, 747 -- 656 BC  \\
26. Dynastie  mit 6 K\"onigen,       \>      Saite, 664 -- 525 BC
\end{tabbing}
Schliesslich gab es noch eine Sp\"ate Periode; sie dauerte von 525 BC bis 332 BC. Sie ist ohne Belang f\"ur die folgenden Betrachtungen.

\section{Chronologie der j\"udischen Geschichte}
Die j\"udische Geschichte beginnt mit der Geburt von Abraham. Sie umfasst, bis zur Zerst\"orung  des ersten Tempels durch den babylonischen  K\"onig Nebuchadrezzar, eine Zeitepoche von ca.\,1500 Jahren. Die Zerst\"orung fand im 18.\,Regierungsjahr von Nebuchadrezzar statt, was in ãTerms of  the Standard Babylonian accession-year SystemÒ dem Jahr 586 BC entspricht. In diesem Jahr wurde  auch die Restbev\"olkerung des jud\"aischen K\"onigreichs in die 70 j\"ahrige Gefangenschaft deportiert. 586 BC ist bis heute die einzige gesicherte Zeitmarke in der j\"udischen Chronologie. Ausgehend von dieser Zeitmarke werden wir im Folgenden versuchen,  r\"uckw\"arts  rechnend, eine widerspruchsfreie  Chronologie bis und mit Moses zu entwickeln. Wir beginnen mit den mehr oder weniger gut bekannten Regierungsdauern der jud\"aaschen K\"onige von Zedekiah bis und mit Solomon, so wie sie Finegan aus der Chronik des Eusebius entnommen hat (Tab.\,1 und 2). Die Addition der insgesamt 21 Regierungszeiten ergibt f\"ur das K\"onigreich Jud\"aa eine Dauer von 446 Jahren. Solomon regierte nach Finegan 40 Jahre, d.h.\,von 1032 BC bis 992 BC. Im 4.\,Jahr seiner Regierungszeit, d.h.\,1028 BC begann er mit dem Bau des ersten Tempels in der j\"udischen Geschichte. Wir sind uns v\"ollig dar\"uber im Klaren, dass diese fr\"uhen Daten nicht mehr ins Konzept der heute viel diskutieren Versuche der Verj\"ungung passen. Wir haben bisher aber keinen stichhaltigen Grund gefunden, warum z.B.\,der Baubeginn und die Regierungszeit von Solomon bis um 100 Jahren reduziert werden sollte. Wir bleiben bei Finegan, der sich auf die Chronologie des Eusebius abst\"utzt und dem Alten Testament. Danach wurde Saul 1112 BC zum 1.\,K\"onig  \"uber die 12 israelitischen  biblischen St\"amme erkoren. Er regierte zusammen mit Samuel w\"ahrend 40 Jahren. Nach Saul folgte David, der 40 Jahre regierte. Solomon herrschte als letzter  K\"onig \"uber das vereinigte Reich der 12 St\"amme. Nach seinem Tod setzten Machtk\"ampfe ein, die mit der Teilung in das s\"udliche K\"onigreich Jud\"aa und das n\"ordliche K\"onigreich Israel endeten. Letzteres existierte nur von 992 BC bis 722 BC. Die resultierende Chronologie ist zu kurz f\"ur  den geplanten Vergleich mit der \"agyptischen Chronologie. Deshalb wenden wir uns wiederum der Chronologie des s\"udlichen Reiches zu. Als Ausgangspunkt w\"ahlen wir jetzt nicht 586 BC sondern beginnen mit Moses und der nachfolgenden Richterzeit. Nach  Finegan war Moses 80 Jahre alt als er sein Volk aus der Sklaverei in \"Agypten erl\"oste und auf eine 40--j\"ahrige Wanderung durch die W\"uste in Richtung des gelobten Landes f\"uhrte. Der Auszug aus \"Agypten, auch als Exodus bezeichnet, muss unter so dramatischen klimatischen Bedingungen statt gefunden haben, dass das Ereignis auch heute noch ''in aller Leute Mund ist''. Zahllose Versuche dieses Datum aus den biblischen Schriften zu gewinnen sind bis heute fehlgeschlagen. Die neusten Sch\"atzungen liegen irgendwo zwischen dem 13.\,und 17.\,Jahrhundert BC.  Nach Moses Tod \"ubernahm Josuha die F\"uhrung w\"ahrend 28 Jahren. W\"ahrend dieser Zeit  begann die noch von Moses geplante Invasion in das gelobte Land namens Kanaan. Sie eroberten im Laufe der Zeit die meisten K\"onigreiche, die damals mit ihren St\"adten und L\"andereien dieses Land besiedelten. Nicht immer waren ihre Kriegsz\"uge erfolgreich. Das wurde zwar im jeweiligen Lagebericht in der Regel auch korrekt vermerkt, meistens aber auch etwas sch\"ongeredet,  wie folgender Auszug aus dem Buch der Richter zeigt:
\begin{quote}
\noindent Richter 3.\,7--9: \emph{Und die Israeliten taten, was dem Herrn missfiel,  und verga\ss en  den Herrn, ihren Gott, und dienten den Baalen und den Ascheren. Da entbrannte der Zorn des  Herrn \"uber Israel, und er verkaufte sie in die Hand Kuschan-Rischatajims, des K\"onigs von Mesopotamien; und so diente Israel dem Kuschan-Rischatajim acht Jahre.}
 
\noindent Richter 3.\,10--11: \emph{Und der Geist des Herrn kam auf ihn (Otniel) und er wurde Richter in Israel und zog aus zum Kampf. Und der Herr gab den K\"onig von Mesopotamien Kuschan-Rischatajims in seine Hand, so dass seine Hand \"uber ihn stark wurde. Da hatte das Land Ruhe vierzig Jahre.}
\end{quote} 
 Auf diese Weise wird im Alten Testament  im Buch der Richter ausf\"uhrlich \"uber die erfolgreichen Taten der 13 einheimischen Richter wie auch \"uber die bitteren 6 Niederlagen mit nachfolgender mehrj\"ahriger Versklavung der Israeliten berichtet (Fig.\,2). Die Fremdherrschaften sind in Fig.\,2 schwarz markiert. Es sind dies die bereits erw\"ahnten 8 Jahre unter den Mesopotamiern, 18 Jahre unter den Moabiter, 20 Jahre unter den  Kanaaniter, 7 Jahre unter den Midianiter, 18 Jahre unter den Philister und Ammoniter, sowie 40 Jahre unter den Philister. Aufsummiert ergibt das 111 Jahre  Fremdherrschaft. Die Richter (inkl.\,Elon, 10 Jahre, den  Finegan weggelassen  hat) sorgten zwischendurch w\"ahrend 338 Jahren f\"ur Krieg und Frieden bis zur Gr\"undung des vereinigten K\"onigreichs durch Samuel und Saul.  Als letzter K\"onig herrschte Solomon \"uber das vereinigte Reich der 12 St\"amme von 1032 BC bis zu seinem Tod  992 BC. Bei Solomon sind wir auf ein Problem gestossen: Solomon hat zwar nach Finegan auch 40 Jahre regiert, gez\"ahlt werden oft aber nur 4 Jahre, weil der Begin des Baus des Tempels  (1028 BC) meistens als Z\"ahlbeginn verwendet wird. Nach seinem Ableben setzten Machtk\"ampfe ein, die mit der Teilung in das s\"udliche K\"onigreich Jud\"aa und das n\"ordliche K\"onigreich Israel endeten. Letzteres existierte von 992 BC bis 722 BC. F\"ur den geplanten Vergleich mit der \"agyptischen Chronologie st\"utzen wir uns  im Folgenden auf die  l\"angere Chronologie der jud\"aischen K\"onige, beginnend mit Rehoboam 992 BC und endend 586 BC mit K\"onig Zedekiah (Tab.\,2) und auf die in Tab.\,1 dargestellten Richterzeiten.

\subsection{Das Exodus Problem}
Die Frage, der wir uns jetzt zuwenden wollen, lautet: Ist es m\"oglich mit Hilfe der beiden
bisher skizzierten Chronologien ein eindeutiges Alter f\"ur den im Alten Testament 
ausf\"uhrlich beschriebenen Exodus zu bestimmen? Um diese Frage ist in j\"ungster Zeit erneut
ein heftiger Streit zwischen den Verfechter der konservativen Hypothesen (das Alte Testament ist unfehlbar) und den Biblischen Minimalisten (das Alte Testament ist ein M\"archenbuch) entbrannt, ohne bisher zu einem Konsens zu gelangen. Zuerst vergleichen wir die absolute Regierungsdauer der in Tab.\,2  aufgef\"uhrten israelitischen Regenten von Moses bis Solomon mit der von  Clayton zusammen gestellten Liste der Pharaos  aus dem gleichen Zeitbereich. In I K\"onige 6, 1 steht, dass Solomon 480 Jahre nach dem Auszug der Israeliten aus \"Agypten, mit dem Bau des Tempels im 4.\,Jahr seiner Regentschaft begann. Nach Finegan w\"are das 1028 BC (Tab.\,2). Wenn wir die 480 Jahre hinzu addieren, finden wir, dass der Exodus um 1508 BC stattgefunden h\"atte. Dieses Datum ist aber wenig wahrscheinlich. Weder in der \"agyptischen noch in der israelitischen Chronologie  findet man um diese Zeit irgendwelche Anhaltspunkte f\"ur den im Alten Testament detailliert  beschriebenen Exodus , es sei denn man identifiziere die Israelis mit den Hyksos, die die \"Agypter etwa um diese Zeit von Osten herkommend, bedr\"angten. Hier stimmt aber die Richtung der Invasoren nicht: Die Israeli wollten raus, die Hyksos  offenbar rein nach \"Agypten! Und das erst noch w\"ahrend der 18.\,Dynastie zu Beginn des  Neuen Reichs. Damit ist auch schon gesagt, dass der Exodus vor Beginn der 18.\,Dynastie, d.h. mindestens 300 Jahre vor der Regierungszeit des Ramses II stattfand. Dieser wohl ber\"uhmteste Pharao regierte von 1279 bis 1212 BC. Sein Nachfolger war Merneptah, sein 13.\,Sohn, der von 1212 bis 1202 BC regierte. Seine Existenz ist nach Clayton [1] dokumentiert durch 3 ber\"uhmte Inschriften: Mehr als 80 Zeilen an einer Wand im Tempel des Amun in Karnak, eine grosse Stele mit 35 Zeilen von Athribis im Delta und die 1896 AD von Flinders Petrie in MernepathÕs Totentempel in  Theben gefundene grosse Siegesstele mit 28 Zeilen. Alle drei befassen sich mit den zahlreichen milit\"arischen Feldz\"ugen dieses Herrschers. MernepathÕs grosse Siegesstele tr\"agt ein Datum (Sommer 1207 BC.) und unter anderem auch folgende Inschrift: \emph{Israel is devastated, her seed  is no more, Palestine has become a widow for Egypt.} Dieser kurze von \"agyptischer Seite erstmalige Hinweis auf eine kriegerische Auseinandersetzung mit den Israeliten wird heute mehrheitlich als Beweis f\"ur den Exodus betrachtet. Dabei wird \"ubersehen, dass nach der  Niederlage die Israeliten offenbar nicht davonrennen konnten, sondern nach der 5.\,Niederlage  als Sklaven f\"ur Ramses II (1279 --1212  BC) w\"ahrend 18 Jahren  und nach der 6.\,Niederlage  w\"ahrend 40 Jahren Frondienst leisten mussten. Die Rolle von Merneptah  (1212 --1202 BC) ist umstritten. Da er nur 10 Jahre regierte, kann er nicht w\"ahrend 18 und 40 Jahren die Arbeiten der Sklaven  \"uberwacht haben. Dasselbe gilt auch f\"ur die folgenden Regenten der 19.\,Dynastie. Viel wahrscheinlicher ist, dass Ramses II f\"ur beide Ereignisse verantwortlich war, f\"ur die w\"ahrend seiner langen Regierungszeit von angeblich 67 Jahren hinreichend Zeit vorhanden war. Rames II war ein \"uberaus kriegerischer Despot, der rund ein Dutzend erfolgreiche Kriegsz\"uge in den umliegenden L\"andern gef\"uhrt hat. Von seinem Vater,  Ramses I wird gesagt [1], dass er nicht von k\"oniglichem Blut sei, sondern aus der Gegend um Avaris im nord-\"ostlichen Delta stamme. Avaris wurde die Hauptstatt der Hyksos nach der vor 400 Jahren gegl\"uckten Invasion in den n\"ordlichen Teil des Nildeltas [1]. Das ist eine interessante Aussage, denn sie k\"onnte m\"oglicherweise den Schl\"ussel zur L\"osung eines R\"atsel sein, das eine rationale Erkl\"arung f\"ur das Exodus Problem bisher verhindert  hat. Das Problem ist folgendes: Wenn Ramses II wirklich von 1279 BC bis 1212 BC und Merneptah, sein 13.\,Sohn, anschliessend bis 1202 BC  regiert haben, dann ergibt sich aus j\"udischer Sicht folgendes Szenario: Um 1261 BC erste Niederlage gegen die Philister mit anschliessender Sklaverei w\"ahrend 18 Jahren. Irgendwann sp\"ater zweite Niederlage wiederum gegen die Philister und Einsatz als Sklaven w\"ahrend 40 Jahren. Es kann nicht unter Mernepthah gewesen sein, weil dieser die Siegess\"aule, auf der vor allem die Taten von Ramses II verherrlicht werden, erst 1212 BC aufstellen liess. Wann der zweite j\"udische Frondienst beendet wurde, ist ebenfalls nicht mit Sicherheit bestimmbar. Nach der j\"udischen Chronologie k\"onnte es 1172 BC gewesen sein, nach der \"agyptischen Chronologie deutlich, n\"amlich rund 30 Jahre fr\"uher, identisch z.B.\,mit dem Ende der Regierungszeit von Merneptah. F\"ur die Israeliten waren Ramses II und Merneptah und der gesamte Clan der 19.\,Dynastie offensichtlich  keine \"Agypter, sondern assimilierte Invasoren, die sich seit rund 400 Jahren im Nildelta festgesetzt hatten und dort mit eigenen Leuten  K\"onigsfolgen und Dynastien gegr\"undet hatten. F\"ur uns gen\"ugt es zu wissen, dass der Exodus keinesfalls in diesem  Zeitbereich stattgefunden hat. Die \"uberlebenden Israeliten wurden hier entweder von Ramses II oder Mernepthah nach der 40 j\"ahrigen Gefangenschaft einfach nach Hause entlassen. Zu vieles spricht daf\"ur, dass sich der Exodus nicht im Zeitbereich der 19.\,Dynastie, sondern einige hundert Jahre fr\"uher abgespielt hat. Wann genau, erz\"ahlt uns die folgende Geschichte.

\subsection{Datierung des Vulkanausbruchs auf Thera (Santorini) mit der  Radiocarbon-Methode}

Die s\"udlichste etwa 120 km n\"ordlich von Kreta gelegene Kykladeninsel Thera, heute als Santorini bezeichnet, ist der \"Uberrest eines grossen Vulkans, der, so vermuteten die Arch\"aologen sp\"atestens in der zweiten H\"alfte des 16.\,Jahrhundert v.\,Chr.\,explodierte. Die Eruption muss verheerende Auswirkungen auf den benachbarten \"ag\"aischen Inseln bis hin zu Kreta und Rhodos gehabt haben. Grosse Mengen von Bimsstein und Asche wurden hoch in die Atmosph\"are geschleudert und riesige Seebebenwellen (Tsunamis) \"uberschwemmten und zerst\"orten die zugewandten K\"ustenregionen im weiten Umkreis. Selbst das globale Klima wurde w\"ahrend Jahren nachhaltig beeinflusst. Auf den \"Uberresten dieser Insel entdeckte der griechische Arch\"aologe Marinatos 1967 in der N\"ahe von Akrotiri, dem heutigen Hauptort  auf dieser Insel, eine minoische Stadt mit mehrst\"ockigen Geb\"auden, die begraben unter  dicken Asche- und Bimssteinschichten zum Teil noch erstaunlich gut erhalten sind. Der genaue Zeitpunkt dieser, die Geschichte der Mittelbronzezeit zweifellos nachhaltig beeinflussende  Katastrophe, war bis vor kurzem umstritten. W\"ahrend die meisten ArchŠologen bis vor kurzem an der zweiten H\"alfte des16.\,Jahrhunderts v.Chr.\,festhielten, sprachen Messungen an S\"aurepeaks in Eisborkernen aus Gr\"onland, Frostsch\"aden an Jahrringen von Eichen in Irland,  Borstenkiefer in Kalifornien und immer pr\"aziseren C14-Datierungen an kurz\-le\.bi\-gen organischen Material aus dem Aschedepot f\"ur ein mindestens um 100 Jahre \"alteres Datum. Bis vor kurzem \"uberschattete ein Sch\"onheitsfehler die C14-Daten. Trotz grossen Anstrengungen und immer  aufwendiger Auswerteverfahren gelang es nicht die Genauigkeit des 2$\sigma$-Bereichs der Datierungen  besser als auf 1663 --1599 BC zu bestimmen. Der Grund: Ein Plateau in der Kalibrierkurve, die bei der Umrechnung der Messdaten in das wahre Alter verwendet werden muss. Dank einem zufŠllig entdecktem Ast eines Olivenbaumes, der am oberen Kraterrand wuchs und w\"ahrend der Eruption brennend von einer rund 60 Meter dicken  Bimsteinschicht zugedeckt wurde, konnte dieses Problem mit der sogenannten Wiggle-Matching Technik vermieden werden [7]. Dieses Auswerte-Verfahren lieferte folgenden best m\"oglichen neuen Wert f\"ur die Eruption des Thera Vulkans. Er lautet:

                            \centerline{{\bf 1627--1600 BC} (2$\sigma$, 95\% confidence limit).}
                             
\noindent So weit so gut. Wie k\"onnen wir aber beweisen, dass der Exodus zeitgleich mit der Eruption statt gefunden hat? Nun beweisen im streng mathematischen Sinn k\"onnen wir es nat\"urlich auch nicht. Wir k\"onnen nur die Indizien zusammentragen, die f\"ur eine Koinzidenz sprechen. 

Der Papyrus Ipuver [8] gibt dazu wichtige Aufschl\"usse. Der  Papyrus mag zur Zeit der 19.\,Dinastie geschrieben worden sein. Der Text bezieht sich aber auf vergangene chaotische Perioden. Widerholt ist auf  \"Ahnlichkeiten zwischen dem Text von Ipuver und dem Bericht von Exodus im Alten Testament hingewiesen worden. Fast alle H\"auser wurden zerst\"ort. Es gab kaum ein Haus ohne einen Toten. Dies weist auf ein Erdbeben gr\"ossten Ausmasses hin. Dabei wurde die Oberschicht, die in Steinh\"ausern wohnte, weit st\*arker betroffen als Sklaven, die in Strohh\"utten hausten. Die \"Armsten wurden reich. Dies erkl\"art, warum die  Israeliten das goldene Kalb giessen konnten. Die n\"achtliche Feuers\"aule und die Rauchs\"aule tags\"uber, welche den Israeliten den Weg wies, d\*urfte von einem weiteren Vulkan stammen, der in Zusammenhang mit dieser Naturkatastrophe ausgebrochen war. 

Im Exodus wird insistiert, dass die Katastrophen vorausgesagt werden konnten. Praktisch sind nur Bewegungen im planetaren Raum voraussagbar. Dies weist auf die M\"oglichkeit hin, dass ein astronomisches Objekt der Ausl\"oser der Katastrophe war.

Eine Naturkatastrophe dieses Ausmasses passt nicht in die Regierungszeit von Ramses II und seiner S\"ohne, die eine Bl\"utezeit der \"agyptischen Geschichte darstellt. Dass der Exodus oft in diese Zeit verlegt wurde, ist mit einer Stele von Merneptah begr\"undet, auf welcher die Isrealiten erstmals erw\"ahnt wurden in Zusammenhang mit zwei Niederlagen.  Wie erw\"ahnt ist anzunehmen, dass sich das auf viel sp\"atere Ereignisse bezieht. 

Der Angelpunkt der hier vorgeschlagenen Chronologie ist die Annahme, dass der Ausbruch von Thera (Santorini) das Naturereignis war, welches den Exodus erm\"oglichte. Die Folge war eine chaotische Zwischenzeit, w\"ahrend der \*Agypten  durch die Hyksos erobert wurde (siehe Ipuver). Deren Herrschaft wurde durch Einwanderungen gefestigt, sowie durch ein durchaus \"agyptisches Verhalten.

\subsection{Schlussfolgerungen}
Da der Ausbruch von Thera den Exodus bewirkte, ist dessen Zeitpunkt durch die C14 Datierung eindeutig auf den Anfang des 17.\,Jahrhunderts festgelegt. Die Katastrophe hatte auf  \"agyptischer Seite den Verlust der Souver\"anit\"at zur Folge. Die Israelis wurden befreit und eroberten ein eigenes Land. In diesen Kriegen gab es auch Niederlagen mit Perioden der Fremdherrschaft, die in Fig.\,2 schwarz gekennzeichnet sind. Die Ber\"ucksichtigung dieser Zeiten auf beiden Seiten f\"uhrt auf eine konsistente Chronologie.

\subsection*{Quellen}
\begin{enumerate}

\item{Peter S. Clayton, {\it Chronicle of the Pharaohs,} Thames and Hudson, 1982. }
\item{{\it Altes Testament } nach der deutschen \"Ubersetzung D.\,Martin Luthers,  W\"urttemberg Bibel, Stuttgart.}
\item{Jack Finegan, {\it Handbook of Biblical Chronology,} Princeton University Press, 1964.}
\item{Manetho, Fragmente seines Buches {\it Geschichte \"Agyptens.} }
\item{Flavius Josephus, {J\"udische Altert\"umer,} I.\,Band, Buch 1--10, Fourier Verlag, Wiesbaden, 7.\,Auflage (1987). }
\item{Georges Bonani, Herbert Hass, Zahi Hawass, Mark Lehner, Shawki Nakhla, John Nolan, Robert Wenke, Willy Woelfli, {\it Radiocarbon dates of old and middle kingdom monuments in Egypt,} Radiocarbon {\bf 43} (3), 1297--1320  (2001).}
\item{Walter L.\,Friedrich, Bernd Kromer, Michael Friedrich, Jan Heinemeier, Tom Pfeiffer, Sahra Talamo, {\it Santorini Eruption Radiocarbon Dated to 1627--1600 B.C.,} Science {\bf 312}, 548 (2006).}
\item{{\it Ipuver,} Papyrus Leiden 334,  http://nefertiti.iwebland.com/texts/ipuwer.htm }.

\end{enumerate}

\newpage
\vspace*{4cm}
\begin{center}
\begin{tabular}{|| l | r | r  | l || }\hline\hline
K\"onig&Dauer &Periode&\hspace{1cm}Bemerkungen\\ \hline
Solomon&36&1028--992& incl.\,Zeit vor Tempelbau:\\ 
Rehoboam&17&992--975&\hspace*{2.75cm} 1032--992\\
Abijam&3&975--972&\\
Asa&41&972--931&\\
Jehoshapht&25&931--906&\\
Jehoram&8&906--898&\\
Ahaziah&1&898--897&\\
Athaliah&7&897--890&\\
Joash&40&890--850&\\
Amaziah&29&850--821&\\
Azariah&52&821--769&\\
Jotham&16&769--753&\\
Ahoz&16&753--737&\\
Hezekiah&29&737--708&\\
Manasseh&55&708--653&\\
Amon&12&653--641&\\
Josiah&32&641--609&\\
Jehoahaz&1&609--608&drei Monate\\
Jehoiakim&11&608--597&\\
Jehoiachin&&&drei Monate\vspace*{-0.27cm}\\ 
&{\large \} }11\vspace*{-0.27cm}&&\\
Zedekiah&&597--586&\\ \hline
\multicolumn{4}{|| l ||}{Seit Beginn des Tempelbaus: $1028-586=442$ Jahre}\\ \hline\hline
\end{tabular}
\end{center}
\centerline{\bf Tab.\,1: Israelische Geschichte nach Solomon aufgrund von Finegan und Altem Testament}

\newpage
\vspace*{2cm}
\begin{center}
\begin{tabular}{|| l | r | l  || }\hline\hline
Moses$^1$ & 120 & 1708 -- 1588 BC  \\
Josuha & 27 & 1588 -- 1561 \\
Aran - Naharim & 8 & 1561 -- 1533  \\
Othniel  & 40 & 1553 -- 1513  \\
Moab & 18 & 1513 -- 1495  \\
Ehud u. Shamgar  & 80 & 1495 -- 1415  \\
Canaaniter & 20 & 1415 -- 1395  \\
Deborah u. Barak & 40 & 1395 -- 1355  \\
Midianiter & 7 & 1355 -- 1348  \\
Gideon & 40 & 1348 -- 1308  \\
Abimelech & 3 & 1308 -- 1305  \\
Tola & 22 & 1305 -- 1283  \\
Jair & 22 & 1283 -- 1261 \\
Philister u. Amoniter & 18 & 1261 -- 1243  \\
Jephthah & 6 & 1243 -- 1237  \\
Ibzan & 7 & 1237 -- 1230  \\
Elon & 10 & 1230 -- 1220  \\
Abdon & 8  & 1220 -- 1212  \\
Philister u. Amelekiter & 40 & 1212 -- 1172  \\
Samson & 20 & 1172 -- 1152  \\
Eli & 40 & 1152 -- 1112  \\
Samuel u. Saul & 40 & 1112 -- 1072  \\
David & 40 & 1072 -- 1032  \\
Solomon$^2$ & 40 & 1032 -- \hspace{0.9 mm} 992 \\ \hline

K\"onigreich Judah$^3$&406 & \hspace{0.08cm} 992 -- \hspace{0.08cm} 586 BC \\ 

K\"onigreich Israel&270& \hspace{0.08cm}  992 -- \hspace{0.08cm} 722 BC  \\   \hline
 \multicolumn{3}{|| l ||}{Exodus  $ \rightarrow$ Tempelbau:  600 Jahre}\\ 
 \multicolumn{3}{|| l ||}{Tempelbau  $ \rightarrow$ 1.\,Zerst\"orung durch}\\
 \multicolumn{3}{|| l ||}{ \hspace*{2.2cm} Babylonier: 442 Jahre}\\
 \multicolumn{3}{|| l ||}{2.\,Zerst\"orung durch R\"omer: 70 AD}\\ \hline
  \multicolumn{3}{|| l ||}{1) Exodus: 1627--1600 BC}\\ 
   \multicolumn{3}{|| l ||}{2)  Beginn des Tempelbaus: 1028 BC}\\
\multicolumn{3}{|| l ||}{3)  1.Zerst\"orung des Tempels: 586 BC} \\  
\hline\hline
\end{tabular}
\end{center}
\centerline{\bf Tab.\,2: Israelische Geschichte vor Solomon aufgrund von Finegan und Altem Testament}

\newpage
\setlength{\topmargin}{1cm}
\setlength{\textheight}{25cm}
\setlength{\oddsidemargin}{0.3cm}
\setlength{\textwidth}{16cm}

\setlength{\unitlength}{0.008cm}

\vspace*{0.1cm}
\hspace*{1cm}
\begin{sideways}

\begin{picture}(3000,1900)(-350,300)
\put(0,2200){\line(1,0){3000}}
\put(0,1900){\line(1,0){3000}}
\put(0,1600){\line(1,0){3000}}
\put(0,1400){\line(1,0){3000}}
\put(0,1200){\line(1,0){3000}}
\put(0,900){\line(1,0){3000}}

\put(0,2200){\line(0,-1){600}}
\put(0,1580){\makebox(0,0)[tl]{\large \bf \"AGYPTEN}}
\put(252.5,2050){\makebox(0,0){\shortstack{Altes Reich\\2686--2181}}}
\put(252.5,1750){\makebox(0,0){\shortstack{Dyn.\,3,4,5,6}}}

\put(505,2200){\line(0,-1){600}}
\put(575.5,2050){\makebox(0,0){\shortstack{1.Zw.\\Zeit}}}
\put(575.5,1750){\makebox(0,0){\shortstack{Dyn.\\7,8\\9,10}}}

\put(646,2200){\line(0,-1){600}}
\put(775,2050){\makebox(0,0){\shortstack{Mittleres\\Reich\\2040--1782}}}
\put(775,1750){\makebox(0,0){\shortstack{Dyn.\\11,12}}}

\put(904,2200){\line(0,-1){600}}
\put(1010,2050){\makebox(0,0){\shortstack{2.Zw.\\Zeit}}}
\put(1010,1750){\makebox(0,0){\shortstack{Dyn.13\\14,15\\16,17}}}

\put(1116,2200){\line(0,-1){600}}
\put(1366,2050){\makebox(0,0){\shortstack{Neues Reich\\1570--1070}}}
\put(1366,1750){\makebox(0,0){\shortstack{Dyn.\,18,19,20}}}

\put(1616,2200){\line(0,-1){600}}
\put(1888.5,2050){\makebox(0,0){\shortstack{3.\,Zwischen-Zeit\\1069--525}}}
\put(1888.5,1750){\makebox(0,0){\shortstack{High Priests\\Dyn.21,22,\\23,24,25,26}}}

\put(2161,2200){\line(0,-1){600}}
\put(2257.5,2050){\makebox(0,0){\shortstack{Sp\"ate\\Zeit\\525--\\332}}}
\put(2257.5,1750){\makebox(0,0){\shortstack{Dyn.27\\28,29\\30,31}}}

\put(2354,2200){\line(0,-1){600}}
\put(2657.5,1750){\makebox(0,0){\shortstack{Maced Kings\\Ptolemaic Dyn.}}}

\put(0,1280){\makebox(0,0)[tl]{\large \bf ISRAEL}}
\put(550,1200){\vector(0,-1){100}}
\put(530,1160){\makebox(0,0)[tr]{\shortstack[r]{2136\\Abraham's\\Geburt}}}
\put(650,1200){\vector(0,-1){100}}
\put(670,1160){\makebox(0,0)[tl]{\shortstack[l]{2036\\Sodom u.\\Gomorah}}}
\put(978,1200){\line(0,-1){80}}
\put(1098,1200){\line(0,-1){80}}
\put(970,1110){\makebox(0,0)[tl]{\shortstack[l]{Moses}}}
\put(1040,910){\makebox(0,0)[br]{\shortstack[r]{Exodus\\1628}}}
\put(1070,955){\makebox(0,0)[bl]{\shortstack[r]{Thera}}}

\put(1654,1200){\line(0,-1){80}}
\put(1694,1200){\line(0,-1){80}}
\put(1640,1110){\makebox(0,0)[tl]{\shortstack[l]{Solomon}}}
\put(1658,900){\vector(0,1){100}}
\put(1699,900){\vector(0,1){100}}
\put(1650,900){\makebox(0,0)[br]{\shortstack[r]{1028\\Tempelbau}}}
\put(1705,900){\makebox(0,0)[bl]{\shortstack[l]{987 Pl\"unde--\\rung d.Tempels}}}
\put(2110,900){\makebox(0,0)[bl]{\shortstack[l]{586\\1. Zerst\"orung\\des Tempels}}}

\put(2764,1200){\vector(0,-1){100}}
\put(2780,1160){\makebox(0,0)[tl]{\shortstack[l]{78 AD\\Vespasian$\dagger$\\Vesuv}}}

\put(1054,1600){\line(0,-1){700}}
\put(1062,1600){\line(0,-1){700}}
\put(2104,1600){\line(0,-1){700}}
\put(2096,1600){\line(0,-1){700}}
\put(2682,1600){\line(0,-1){700}}
\put(2690,1600){\line(0,-1){700}}

\put(86,1400){\line(0,1){50}}
\put(186,1400){\line(0,1){70}}
\multiput(286,1400)(100,0){4}{\line(0,1){50}}
\put(686,1400){\line(0,1){100}}
\multiput(786,1400)(100,0){4}{\line(0,1){50}}
\put(1186,1400){\line(0,1){70}}
\multiput(1286,1400)(100,0){4}{\line(0,1){50}}
\put(1686,1400){\line(0,1){100}}
\multiput(1786,1400)(100,0){4}{\line(0,1){50}}
\put(2186,1400){\line(0,1){70}}
\multiput(2286,1400)(100,0){4}{\line(0,1){50}}
\put(2686,1400){\line(0,1){100}}
\multiput(2786,1400)(100,0){3}{\line(0,1){50}}
\put(186,1350){\makebox(0,0){2500}}
\put(686,1350){\makebox(0,0){2000}}
\put(1186,1350){\makebox(0,0){1500}}
\put(1686,1350){\makebox(0,0){1000}}
\put(2186,1350){\makebox(0,0){500}}
\put(2676,1350){\makebox(0,0){BC\hspace{2mm}   0   \hspace*{2mm} AD}}

\put(1400,750){\makebox(0,0){\bf Fig.\,1:  \"Ubersicht \"uber die \"Agyptische und Isrealische Chronošogie}}

\end{picture}

\end{sideways}

\newpage

\setlength{\unitlength}{0.03cm}
\vspace*{0.1cm}
\hspace*{1cm}
\begin{sideways}

\begin{picture}(800,500)(-100,0)

\put(0,530){\line(1,0){800}}
\put(0,490){\line(1,0){800}}
\put(0,450){\line(1,0){800}}
\put(0,370){\line(1,0){800}}
\put(0,300){\line(1,0){800}}
\put(0,260){\line(1,0){800}}
\put(0,160){\line(1,0){800}}

\put(37,370){\line(0,1){120}}
\put(130,450){\line(0,1){80}}
\multiput(130,370)(0,20){6}{\line(0,1){15}}
\put(154,370){\line(0,1){80}}
\put(176,370){\line(0,1){80}}
\put(407,370){\line(0,1){120}}
\put(422,370){\line(0,1){80}}
\put(488,370){\line(0,1){80}}
\put(498,370){\line(0,1){80}}
\put(515,370){\line(0,1){120}}
\put(630,370){\line(0,1){160}}
\put(755,450){\line(0,1){40}}

\put(0,495){\makebox(0,0)[bl]{\shortstack[l]{Zweite Zwischenzeit\\1782--1570 (55)}}}
\put(380,508){\makebox(0,0){\shortstack{Neues Reich\\1650 (70) -- 1070}}}
\put(800,500){\makebox(0,0)[br]{\shortstack[r]{Dritte Zwischenzeit\\1069--525}}}
\put(83,470){\makebox(0,0){\shortstack{17. Dynastie\\1663--1570}}}
\put(268,470){\makebox(0,0){\shortstack{18. Dynastie\\1570--1293}}}
\put(464,470){\makebox(0,0){\shortstack{19. Dynastie\\1293--1185}}}
\put(572,470){\makebox(0,0){\shortstack{20. Dynastie\\1185--1070}}}
\put(692,470){\makebox(0,0){\shortstack{11. Dyn. 1065--945\\{\small High Priests 1080--945}}}}
\put(83,410){\makebox(0,0){\shortstack{15.16.Dynastie\\(Hyksos)\\1663--1570}}}
\put(138,440){\makebox(0,0)[tl]{\begin{sideways} Ahmose I\end{sideways} }}
\put(160,449){\makebox(0,0)[tl]{\begin{sideways} Amenhotep I\end{sideways} }}
\put(409,448){\makebox(0,0)[tl]{\begin{sideways} Ram.I,Seti I \end{sideways} }}
\put(455,410){\makebox(0,0){\shortstack{Rames--\\ses II}}}
\put(488,443){\makebox(0,0)[tl]{\begin{sideways} Merneptah \end{sideways} }}
\put(573,410){\makebox(0,0){\shortstack{Setnakhte (3)\\Ramesses III--X}}}

\put(0,300){\line(0,1){22}}
\multiput(10,300)(10,0){4}{\line(0,1){15}}
\put(50,300){\line(0,1){18}}
\multiput(60,300)(10,0){4}{\line(0,1){15}}
\put(100,300){\line(0,1){22}}
\multiput(110,300)(10,0){4}{\line(0,1){15}}
\put(150,300){\line(0,1){18}}
\multiput(160,300)(10,0){4}{\line(0,1){15}}
\put(200,300){\line(0,1){22}}
\multiput(210,300)(10,0){4}{\line(0,1){15}}
\put(250,300){\line(0,1){18}}
\multiput(260,300)(10,0){4}{\line(0,1){15}}
\put(300,300){\line(0,1){22}}
\multiput(310,300)(10,0){4}{\line(0,1){15}}
\put(350,300){\line(0,1){18}}
\multiput(360,300)(10,0){4}{\line(0,1){15}}
\put(400,300){\line(0,1){22}}
\multiput(410,300)(10,0){4}{\line(0,1){15}}
\put(450,300){\line(0,1){18}}
\multiput(460,300)(10,0){4}{\line(0,1){15}}
\put(500,300){\line(0,1){22}}
\multiput(510,300)(10,0){4}{\line(0,1){15}}
\put(550,300){\line(0,1){18}}
\multiput(560,300)(10,0){4}{\line(0,1){15}}
\put(600,300){\line(0,1){22}}
\multiput(610,300)(10,0){4}{\line(0,1){15}}
\put(650,300){\line(0,1){18}}
\multiput(660,300)(10,0){4}{\line(0,1){15}}
\put(700,300){\line(0,1){22}}
\multiput(710,300)(10,0){4}{\line(0,1){15}}
\put(750,300){\line(0,1){18}}
\multiput(760,300)(10,0){4}{\line(0,1){15}}
\put(800,300){\line(0,1){22}}

\put(0,330){\makebox(0,0){\hspace*{8mm}1700 BC}}
\put(100,330){\makebox(0,0){1600}}
\put(200,330){\makebox(0,0){1500}}
\put(300,330){\makebox(0,0){1400}}
\put(400,330){\makebox(0,0){1300}}
\put(500,330){\makebox(0,0){1200}}
\put(600,330){\makebox(0,0){1100}}
\put(700,330){\makebox(0,0){1000}}
\put(800,330){\makebox(0,0){900}}

\put(0,363){\makebox(0,0)[tl]{\large \bf \"AGYPTEN}}
\put(0,286){\makebox(0,0)[tl]{\large \bf ISRAEL}}

\put(112,160){\line(0,1){100}}
\put(139,160){\line(0,1){100}}
\put(147,160){\line(0,1){100}}
\put(187,160){\line(0,1){100}}
\put(205,160){\line(0,1){100}}
\put(285,160){\line(0,1){100}}
\put(305,160){\line(0,1){100}}
\put(345,160){\line(0,1){100}}
\put(352,160){\line(0,1){100}}
\put(392,160){\line(0,1){100}}
\put(395,160){\line(0,1){100}}
\put(417,160){\line(0,1){100}}
\put(439,160){\line(0,1){100}}
\put(457,160){\line(0,1){100}}
\put(463,160){\line(0,1){100}}
\put(470,160){\line(0,1){100}}
\put(480,160){\line(0,1){100}}
\put(488,160){\line(0,1){100}}
\put(528,160){\line(0,1){100}}
\put(548,160){\line(0,1){100}}
\put(588,160){\line(0,1){100}}
\put(628,160){\line(0,1){100}}
\put(668,160){\line(0,1){100}}
\put(708,160){\line(0,1){100}}
\put(728,160){\line(0,1){100}}

\put(52,210){\makebox(0,0){\shortstack{\small Moses\\120}}}
\put(125,210){\makebox(0,0){\shortstack{\small Jesu--\\ha\\27}}}
\put(143,210){\makebox(0,0){\shortstack{\small A\\N}}}
\put(167,210){\makebox(0,0){\shortstack{\small Othniel\\40}}}
\put(196,210){\makebox(0,0){\shortstack{\small Mo-\\ab\\18}}}
\put(245,210){\makebox(0,0){\shortstack{\small Ehud\\80}}}
\put(295,210){\makebox(0,0){\shortstack{\small Can.\\20}}}
\put(325,210){\makebox(0,0){\shortstack{\small Debo-\\rah\\v.B.\\40}}}
\put(348.5,210){\makebox(0,0){\shortstack{\small M\\7}}}
\put(372,210){\makebox(0,0){\shortstack{\small Gideon\\40}}}
\put(393.5,210){\makebox(0,0){\shortstack{\small A}}}
\put(406,210){\makebox(0,0){\shortstack{\small T\\22}}}
\put(428,210){\makebox(0,0){\shortstack{\small J\\22}}}
\put(448,210){\makebox(0,0){\shortstack{\small Ph\\Ap}}}
\put(460,210){\makebox(0,0){\shortstack{\small J\\6}}}
\put(466.5,210){\makebox(0,0){\shortstack{\small I\\7}}}
\put(475,210){\makebox(0,0){\shortstack{\small E\\10}}}
\put(484,210){\makebox(0,0){\shortstack{\small A\\8}}}
\put(508,210){\makebox(0,0){\shortstack{\small Phil.\\Amel.}}}
\put(538,210){\makebox(0,0){\shortstack{\small S\\20}}}
\put(568,210){\makebox(0,0){\shortstack{\small Eli\\40}}}
\put(608,210){\makebox(0,0){\shortstack{\small Sam\\Saul\\40}}}
\put(648,210){\makebox(0,0){\shortstack{\small David\\40}}}
\put(688,210){\makebox(0,0){\shortstack{\small Solo-\\mon\\40}}}
\put(718,210){\makebox(0,0){\shortstack{\small J\\20}}}

\put(139,230){\makebox(0,0)[bl]{\rule{0.24cm}{0.9cm}}}
\put(139,190){\makebox(0,0)[tl]{\rule{0.24cm}{0.9cm}}}
\put(187,230){\makebox(0,0)[bl]{\rule{0.54cm}{0.9cm}}}
\put(187,190){\makebox(0,0)[tl]{\rule{0.54cm}{0.9cm}}}
\put(285,230){\makebox(0,0)[bl]{\rule{0.6cm}{0.9cm}}}
\put(285,190){\makebox(0,0)[tl]{\rule{0.6cm}{0.9cm}}}
\put(345,230){\makebox(0,0)[bl]{\rule{0.21cm}{0.9cm}}}
\put(345,190){\makebox(0,0)[tl]{\rule{0.21cm}{0.9cm}}}
\put(439,230){\makebox(0,0)[bl]{\rule{0.54cm}{0.9cm}}}
\put(439,190){\makebox(0,0)[tl]{\rule{0.54cm}{0.9cm}}}
\put(488,230){\makebox(0,0)[bl]{\rule{1.2cm}{0.9cm}}}
\put(488,190){\makebox(0,0)[tl]{\rule{1.2cm}{0.9cm}}}

\put(115,135){\vector(0,1){23}}
\put(115,127){\makebox(0,0){\shortstack{ Jericho}}}
\put(70,87){\vector(0,1){71}}
\put(672,87){\vector(0,1){71}}
\put(70,102){\vector(1,0){600}}
\put(371,92){\makebox(0,0){\shortstack{ 600 Jahre}}}
\put(70,75){\makebox(0,0){\shortstack{ Exodus 1627--1600 BC}}}
\put(672,75){\makebox(0,0){\shortstack{ Tempelbau\\1028 BC}}}

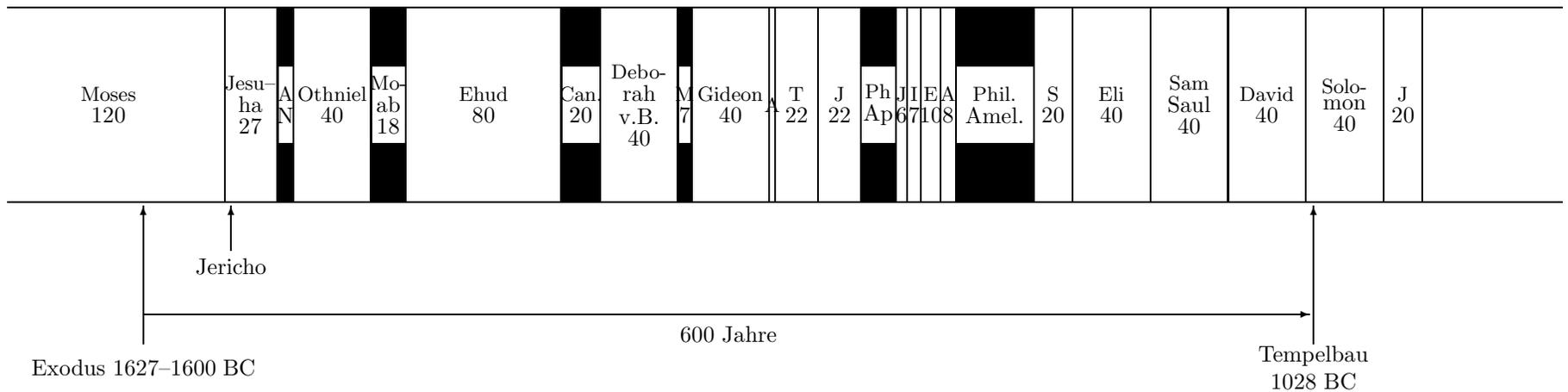
\put(0,25){\makebox(0,0)[tl]{\shortstack[l]{\bf Fig.2:  \"Agyptische und Israelische Chronologie im Bereich der Richterzeit nach Finegan, Altem Testament \\ \bf \hspace*{1.3cm} und Clayton.}}}

\end{picture}

\end{sideways}

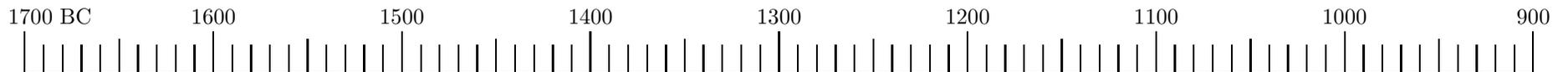
\end{document}